\documentclass[aps,prl,10pt,twocolumn,notitlepage,nofootinbib,longbibliography,showpacs]{revtex4-1}

\usepackage{amsmath,bm}
\usepackage{amsfonts}%
\usepackage{amsmath}%
\usepackage{amssymb}%
\usepackage{bm}%
\usepackage{bbold}%
\usepackage{latexsym}%
\usepackage{graphicx}
\usepackage{hyperref}

\newcommand{\bmath}{\begin{displaymath}}
\newcommand{\emath}{\end{displaymath}}
\newcommand{\bite}{\begin{itemize}}
\newcommand{\eite}{\end{itemize}}

\newcommand{\eps}{\varepsilon}

\newcommand{\half}{{\textstyle \frac{1}{2}}}
\newcommand{\E}{{\mathcal{E}}}

\newcommand{\bel}[1]{\begin{equation}\label{#1}}
\newcommand{\bal}[1]{\begin{eqnarray}\label{#1}}
\newcommand{\ee}{\end{equation}}
\newcommand{\ea}{\end{eqnarray}}

\newcommand{\tphi}{{\sigma}}

\newcommand{\fig}[1]{fig.~\ref{#1}}

\newcommand{\equ}[1]{eq.(\ref{#1})}
\newcommand{\Equ}[1]{Eq.(\ref{#1})}

\begin{document}

\title{Quantization and Renormalization and the Casimir Energy of a Scalar Field Interacting with a Rotating Ring}
\author{M. Schaden}
\email{mschaden@rutgers.edu}
\homepage{http://www.ncas.rutgers.edu/martin-schaden}
\affiliation{Department of Physics, 
Rutgers, The State University of New Jersey, 
101 Warren Street, Newark, New Jersey - 07102, USA.}
%\author{K. V. Shajesh}
%\email{shajesh@andromeda.rutgers.edu}
%\homepage{http://andromeda.rutgers.edu/~shajesh}
%\affiliation{Ludovici Caddemartiri\rq{}s Lab???}

\date{\today}
%need to use showpacs in displayclass options
%\keywords{} %need to use showkeys in displayclass options

%--------------------------------------------
\begin{abstract}
Effects due to vacuum fluctuations in a semi-classical model of a massless scalar field interacting with a rotating ring are investigated by introducing a collective coordinate for the motion of the background potential. The model is solved for a repulsive periodic $\delta$-distribution background of arbitrary strength. The Casimir energy of this system is calculated in the co-rotating and, by Legendre transformation, in the stationary laboratory frame. The zero-point contribution to the angular momentum in this model is bounded below by $|\ell_{ZP}|\leq\hbar/24$ and the ground state of the entire system thus generally is non-rotating with a positive moment of inertia that decreases only slightly with increasing angular rotation frequency. There is no transfer between the zero-point and classical contributions to the total angular momentum and energy of this system at zero temperature.        
\end{abstract}

%--------------------------------------------
%\tableofcontents
%--------------------------------------------
\maketitle

%%\section{Introduction}
Recently Chernodub observed \cite{Chernodub:2012ry,Chernodub:2012em} that zero-point fluctuations of a scalar field contribute negatively to the moment of inertia and speculated that under favorable circumstances the ground state of some macroscopic devices could be self-rotating and thus realize a quantum time-crystal \cite{Shapere:2012nq,Wilczek:2012tc}. Here the simplest of such models \cite{Chernodub:2012ry,Chernodub:2012em} is examined in greater detail. I extend the original model  to admit less singular interactions and conserve total angular momentum and energy by introducing a collective coordinate \cite{Bohr:1969ws} that describes the rotation dynamically.

Collective coordinates were used by Lord Rayleigh to describe classical vibrations of a droplet. Bohr and Mottelson \cite{Bohr:1969ws,Mottelson:1976sc} introduced them in the context of nuclear physics to represent collective aspects of many-body quantum mechanics. This approach also describes the dynamics of soliton moduli spaces in field theory \cite{Sutcliffe:1993su}. However, these methods do not appear to have been systematically exploited in dynamic Casimir effects. They are capable of resolving long-standing issues in this field and provide a general and rigorous framework for handling vacuum effects arising from dynamical classical systems.          

%\section{The Extended Chernodub Model}
Let us for example consider the relatively simple and instructive example model in which a scalar field on a circle of radius $R$ interacts with an everywhere positive background field (potential) $V(\varphi)\ge 0$ whose position on the circle is referenced by the collective coordinate $\theta(t)$. The Lagrangian for this model is, 
\begin{align}
\label{L}
L(\phi,\dot\phi,\theta,\dot\theta)\!=\!\frac{I}{2 }\dot\theta^2\!+\!\oint_{S_1}\hspace{-0.5em} R d\varphi \half\left(\dot\phi^2\!-\!R^{-2}\phi^{\prime\; 2}\!-\!V(\varphi\!-\!\theta(t))\;\phi^2\right)\ , 
\end{align}
where $I$ is the moment of inertia for the collective coordinate $\theta(t)$. \Equ{L} may be  interpreted as describing quantum fluctuations to quadratic order of a scalar field $\phi(\varphi,t)$ on a circle in the background of a classical soliton solution $V(\varphi)$ located at $\theta(t)$. This only omits to specify the originally highly nonlinear model $V(\phi)$ is the soliton of. All Casimir systems could (and perhaps should) be interpreted in this manner \cite{Graham:2002fi}. The presence of a \lq\lq{}classical\rq\rq{} contribution $I\dot\theta^2/2$ to the rotational energy in this case is due to the motion of the soliton on the circle (or that of a ring) and is not at all surprising. The original Chernodub model of a scalar on a circle subject to a rotating Dirichlet boundary condition \cite{Chernodub:2012ry} is the limit of this extended model for very large $I\gg \hbar R/c$ and a \lq\lq{}thin-wall\rq\rq{} soliton $V(\varphi)$ proportional to a periodic $\delta$-distribution at strong coupling $\lambda\rightarrow\infty$. The main qualitative modification to Chernodub\rq{}s original model thus is the presence of a dynamical collective coordinate giving the location and dynamics of the wall.   

The extended Lagrangian of \equ{L} does not explicitly depend on time and conserves total angular momentum. We are interested in the lowest (vacuum) energy of this model for a given total angular momentum $\ell$. For a quantum time crystal, the energy is minimal at $\ell\neq 0$.

Written in terms of the relative angle $\tphi=\varphi-\theta(t)$, the Lagrangian of \equ{L} reads,
\bel{Lrot}
L(\phi,\dot\phi,\dot\theta)= \frac{I}{2}\dot\theta^2 +\oint_{S_1}\hspace{-0.5em} R d\tphi \half\left((\dot\phi-\dot\theta\phi^{\prime})^2 -R^{-2}\phi^{\prime\; 2} -V(\tphi)\;\phi^2\right)\ ,
\ee 
and $\theta$ is a cyclical coordinate. The canonical momenta are,  
\begin{subequations}
\begin{align}
\pi(\tphi,t)&=\dot\phi(\tphi,t)-\dot\theta \phi^{\prime}(\tphi,t)
\label{pi}\\
\ell&=I\dot\theta-\oint_{S_1}\hspace{-0.5em} R d\tphi\; \pi(\tphi,t)\phi^{\prime}(\tphi,t)\ ,
\label{ell}
\end{align}
\label{Pis}
\end{subequations}
where $\ell$ defined by \equ{ell} is the conserved total angular momentum conjugate to $\theta$.
The Hamiltonian $H_s$ in the ($s$)tationary frame in these coordinates is obtained from \equ{Lrot} as,
\begin{align}
\label{Hstationary}
H_s&=-L+\ell\dot\theta+\oint_{S_1}\hspace{-0.5em} R d\tphi\;\pi(\tphi,t)\dot\phi(\tphi,t)  \nonumber\\
&=\frac{1}{2 I}\left(\ell+\oint_{S_1} \hspace{-0.5em}R d\tphi\; \pi \phi^{\prime}\right)^2\\
&\hspace{2em}+\oint_{S_1} \hspace{-0.5em} R d\tphi\; \half\left(\pi^2 +R^{-2}\phi^{\prime\; 2} +V(\tphi)\;\phi^2\right)\ .\nonumber
\end{align}
The quartic term of $H_s$ is linearized by a Legendre transformation to the energy $H_c$ in the co-rotating frame. With the angular rotation frequency $\Omega\neq\dot\theta$ given as, 
\bel{defOm}
\Omega=\frac{\partial H_s}{\partial \ell}=\frac{1}{I}\left(\ell+\oint_{S_1} R d\tphi \pi(\tphi,t) \phi^{\prime}(\tphi,t)\right)
\ee
we have that (see footnote on p.74 of  \cite{Landau:1980li}), 
\begin{align}
\label{Hc}
H_c(\Omega)&=H_s(\ell(\Omega))-\ell(\Omega) \Omega\\
&\hspace{-3em}=\!-\frac{I}{2}\Omega^2+\!\oint_{S_1}\hspace{-0.5em} R d\tphi \half\left(\pi^2\!+\!\Omega(\pi\phi^\prime\!+\!\phi^\prime\pi)\!+\! R^{-2}\phi^{\prime\; 2}\! +\!V(\tphi)\phi^2\right) ,\nonumber
\end{align}
is  quadratic in fields and momenta.  Canonical quantization of this model proceeds by promoting fields and momenta to operators with equal time commutator,
\bel{quant}
[\phi(\tphi,t), \pi(\tphi^\prime,t)]=\frac{i}{R}\delta_\text{per}(\tphi-\tphi^\prime)\ ,
\ee 
where $\delta_\text{per}(\sigma)=\frac{1}{2\pi}\sum_{n\in\mathbb{Z}} e^{i n \sigma}$ denotes the periodic $\delta$-distribution. With the ordering given in \equ{Hc}, $H_c$ is hermitian and Hamilton\rq{}s equations,
\begin{subequations}
\begin{align}
\dot\phi&=\frac{\delta H_c}{\delta \pi}=\pi+\Omega\phi^\prime
\label{em1}\\
\dot\pi&=-\frac{\delta H_c}{\delta \phi}=\Omega \pi^\prime+R^{-2}\phi^{\prime\prime}-V(\tphi)\phi\ ,
\label{em2}
\end{align}
\label{Heqm}
\end{subequations}
coincide with Heisenberg\rq{}s equations of motion for the operators. Solving for $\pi$ in \equ{em1} and inserting in \equ{em2}  gives the separable equation of motion in the co-rotating frame,
\bel{eqm}
\ddot\phi-2\Omega\dot\phi^\prime+(\Omega^2-R^{-2})\phi^{\prime\prime}+V(\tphi)\phi=0\ ,
\ee
whose general solution is,
\bel{solution1} 
\phi(\tphi,t)= \sum_{\omega_m\ge0}^\infty \ a_m e^{-i \omega_m t} u_m (\tphi) +a^\dagger_m e^{i \omega_m t}\bar u_m (\tphi) \ .
\ee
Upon quantization, the coefficients $a_m$ and $a^\dagger_m$ are interpreted as annihilation and creation operators of quanta in mode $m$. Inserting the general solution of \equ{solution1} in \equ{em1}, the momentum is expressed in terms of operators $a$ and $a^\dagger$ as,
\begin{align}
\label{momentum}
\pi(\tphi,t)=\dot\phi-\Omega\phi^\prime&=\sum_{\omega_m\ge0}^\infty \ a_m e^{-i \omega_m t}(-i\omega_m u_m-\Omega u_m^\prime)\nonumber\\
&\hspace{3em} +a^\dagger_m e^{i \omega_m t}(i\omega_m \bar u_m-\Omega \bar u_m^\prime)\ .
\end{align} 
 \equ{eqm} implies that the mode functions $u_m(\tphi)$  satisfy the homogeneous ODE,
\bel{um}
(\Omega^2-R^{-2})u^{\prime\prime}_m+2 i\omega_m\Omega u^\prime_m +(V(\tphi)-\omega_m^2) u_m=0\ .
\ee
of Bloch waves \cite{Bloch:1929bl}. The frequencies $\omega_m$ at which \equ{um} has non-trivial periodic solutions are discrete. For a finite periodic potential $\infty>V(\tphi)=V(\tphi+2\pi)\ge0$, the frequencies $\omega_m$ and corresponding solutions  $u_m(\tphi)$  of \equ{um} are determined by requiring that,
\bel{bcs}
\left(\begin{smallmatrix} u_m(\tphi)\\u_m^\prime(\tphi)\end{smallmatrix}\right)=\lim_{\eps\rightarrow 0_+}\left(\begin{smallmatrix} u_m(\tphi+2\pi-\epsilon)\\u_m^\prime(\tphi+2\pi-\epsilon)\end{smallmatrix}\right)
\ee 
The frequency spectrum $\{\omega_m\}$ is real and the complex conjugate mode function $\bar u_m(\tphi)$  is a solution of \equ{um} to frequency $-\omega_m$. It thus suffices in \equ{solution1} to sum over non-negative frequencies only.  \Equ{um} is consistent with the normalization conditions,
\begin{align}
\label{norm}
R\oint_{S_1} d\tphi \left[(\omega_m+\omega_n) \bar u_m u_n -2 i \Omega\bar u_m u_n^\prime\right]&=\delta_{mn}\nonumber\\
R\oint_{S_1} d\tphi \left[(\omega_n-\omega_m) u_m u_n -2 i \Omega u_m u_n^\prime\right]&=0,
\end{align}
and their complex conjugates. Mode functions to different frequencies are orthogonal in this sense and \equ{norm} can be satisfied when some frequencies happen to be degenerate. Using \equ{norm} in \equ{solution1} and \equ{momentum}  the annihilation operators are related to field operators as,  
\begin{align}
\label{as}
a_n&=R\oint_{S_1} d\tphi (\phi(\tphi,0) \omega_n + i\pi(\tphi,0) )\bar u_n\ ,\nonumber\\
a^\dagger_n&=R\oint_{S_1} d\tphi (\phi(\tphi,0) \omega_n - i\pi(\tphi,0) )\bar u_n\ .
\end{align}
\Equ{as}, \equ{quant}  and \equ{norm} imply the usual commutation relations,
\bel{can}
[a_m,a_n]=[a_m^\dagger,a_n^\dagger]=0\ ;\ \ [a_m,a_n^\dagger]=\delta_{mn}\ ,
\ee
of creation and annihilation operators. 
Inserting \equ{solution1} and \equ{momentum} in \equ{Hc} and using \equ{norm}, the Hamiltonian $H_c(\Omega)$ of the co-rotating system is seen to be diagonal,
\bel{Hq}
H_c(\Omega)=\half\sum_{\omega_n\ge0} \omega_n(\Omega,R) (a^\dagger_n a_n +a_n a^\dagger_n) \ ,
\ee
and the construction of the Fock-space is analogous to the non-rotating case:  at any given angular frequency $\Omega$, the lowest energy state of the co-rotating system is annihilated by all $a_n$ and has the zero-point energy given by the formal sum,
\bel{zpE}
E_{ZP}(R,\Omega) ="\half \sum_{\omega_n\ge0} \omega_n(R,\Omega)\ \ "\ .
\ee
Since we are interested only in the dependence of this energy on external parameters like the radius $R$ and angular rotation frequency $\Omega$, we may instead compute the finite\footnote{A finite single-particle Casimir energy can be defined only if a certain coefficient of the asymptotic heat kernel expansion of the differential operator vanishes -- for scalar fields on $S_1$ this is the case and the Casimir energy is finite for any positive potential as well as for a Dirichlet condition \cite{Kirsten:2001wz}.}  difference,
\bel{Cas}
\E_c(R,\Omega)= E_{ZP}(R,\Omega) - E_{ZP}(\infty,0)\ ,
\ee  
which one may refer to as the Casimir energy of the co-rotating system. Various methods have been developed to extract the parameter-dependent part of the infinite zero-point energy. Here this is straightforward only if the intermediate regularization of \equ{zpE} respects the symmetries of the \emph{co-rotating} system. Many regularizations, ranging from the insertion of an exponential cutoff in the zero-point sum of \equ{zpE} to generalized zeta-function regularization, to point-splitting, meet this criterion. However, in the latter regularization method, the point-splitting should be invariant under the time-translation symmetry of the co-rotating frame. This is not the same as time-splitting in the lab-frame. The point-splitting regularization otherwise explicitly breaks rotational invariance, which would have to be explicitly restored for the total angular momentum to be conserved. 

%\section{A $\delta$-Distribution Background}
We further restrict our considerations to the example of a periodic $\delta$-distribution background $V(\tphi)=\lambda\delta_\text{per}(\sigma)$. The previous considerations for \emph{finite} potentials are readily adapted to this singular case. \Equ{um} and the boundary conditions of \equ{bcs}  for a periodic $\delta$-distribution potential become,
\begin{subequations}
\begin{align}
&0=(1-\beta^2) u^{\prime\prime}_m-2 i\alpha_m\beta u^\prime_m +\alpha_m^2 u_m
\label{eq}\\
&\text{with} \ \ u_m(0)=u_m(2\pi) 
\label{cond1}\\ 
&\text{and}\ u_m^\prime(0)-u_m^\prime(2\pi)=\frac{\lambda R^2}{1-\beta^2}u_m(0)\ ,
\label{cond2}
\end{align}
\label{singular}
\end{subequations}
where $\alpha_m=\omega_m R/c$ and $\beta=\Omega R/c$ are the dimensionless frequency and rotation speed. \Equ{cond2} ensures that the  discontinuity in the derivative of $u_m$ compensates for the singular potential in \equ{um}.  The mode function satisfying \equ{singular} to the dimensionless frequency $\alpha_m$ is of the form,
\bel{umSing}
u_m(\tphi)\propto(1-e^{-\frac{2\pi i\alpha_m}{1+\beta}}) e^{i\tphi\frac{\alpha_m}{1-\beta}} - (1-e^{\frac{2\pi i\alpha_m}{1-\beta}}) e^{-i\tphi\frac{\alpha_m}{1+\beta}}\ ,
\ee
with  $\alpha =\alpha_m$ a solution to the secular equation,
\bel{secular}
\sin\frac{\pi\alpha}{1-\beta}\sin\frac{\pi\alpha}{1+\beta}=\frac{\lambda R^2}{4\alpha} \sin\frac{2\pi\alpha}{1-\beta^2}\ .
\ee
Note that for $\lambda\sim\infty$ the mode function in \equ{umSing}, satisfies the Dirichlet condition $u_m(0)=0$.

Using the generalized argument principle \cite{VanKampen:1968}, \equ{secular} gives the Casimir energy in the co-rotating frame of a scalar field interacting with a rotating ring by a periodic $\delta$-distribution potential of strength $\lambda$ for any\footnote{The generalized $\zeta$-function techniques of  \cite{Kirsten:2001wz} allow one to numerically obtain this energy for any  well-behaved potential $V(\tphi)$.}  radius $R$ and angular rotation frequency $\Omega$ as the finite integral,
\begin{align}
\label{CasE}
\E_c(R, \beta=\Omega R/c,\lambda )&=\\
&\hspace{-8em}=\frac{\hbar c}{2\pi R}\!\int_0^\infty \hspace{-1em}d\zeta \ln\!\Big(\!1\!-\!\frac{4 \zeta \cosh\left(\frac{2\pi\zeta\beta}{1-\beta^2}\right)\!+\!(\lambda R^2\!-\!2\zeta) e^{-\frac{2\pi\zeta}{1-\beta^2}}}{(2\zeta\!+\!\lambda R^2)e^{\frac{2\pi\zeta}{1-\beta^2}}}\!\Big)\ .\nonumber
\end{align}
One can perform the integral analytically in the limits of vanishing and very strong coupling: $\E_c(R, \beta,\lambda=0)=-\frac{\hbar c}{12 R}$ and $\E_c(R, \beta,\lambda\sim\infty)=-\frac{\hbar c}{48 R}(1-\beta^2)$. For vanishing interaction strength, the frequencies solving \equ{secular} of left and right-moving modes are Doppler-shifted by factors $1\pm \beta$. Their sum and thus the Casimir energy of the co-rotating frame do not depend on $\Omega$. In the limit of very strong interaction strength on the other hand, the spectrum of frequencies solving \equ{secular} is $\{\omega_m=(1-\Omega^2 R^2) \frac{m}{2R}; m\in \mathbb{N}\}$ and the dependence  of the Casimir energy on $\Omega$ is quadratic in the co-rotating frame, 
\bel{casinf}
\E^\infty_c\!=\!-\frac{\hbar c}{48 R}(1-\beta^2)-\frac{I}{2}\Omega^2\!=\!-\frac{\hbar c}{48 R}+\frac{\hbar R\Omega^2}{48 c}-\frac{I}{2}\Omega^2\ .
\ee
The inverse Legendre transform of $\E^\infty_c$ in \equ{casinf} gives the dependence of the ground state energy on the total angular momentum,
\bel{angular}
\ell(\lambda\sim\infty)=-\left.\frac{\partial \E_c}{\partial \Omega}\right|_{\lambda\rightarrow\infty}= \left(I-\frac{\hbar}{24 c} R\right)\Omega\ ,
\ee
as,
\begin{align}
\label{Es}
\E_s &=\E_c(\Omega)+\ell\Omega\\
&\genfrac{}{}{0pt}{1}{-\!\!\!-\!\!\!-\!\!\!-\!\!\!\longrightarrow}{\lambda\rightarrow\infty} -\frac{\hbar c}{48 R}+\frac{\ell^2}{2(I-\frac{\hbar}{24c} R)}=-\frac{\hbar c}{48 R}-\frac{\hbar R}{48 c} \Omega^2+\frac{I}{2} \Omega^2\ .\nonumber
\end{align}
Apart from the classical contribution proportional to $I$, \equ{Es} reproduces the zero-point energy of a scalar field with rotating Dirichlet boundary conditions obtained in \cite{Chernodub:2012ry}. Although the computation of \cite{Chernodub:2012ry} in the stationary frame for general potentials does not conserve energy and angular momentum, our results do agree for a rotating Dirichlet boundary conditions. 
As Chernodub pointed out \cite{Chernodub:2012ry}, \equ{Es} implies that zero-point fluctuations of a scalar field \emph{reduce} the moment of inertia of the device.  However, the classical contribution in general is not negligible and the semi-classical treatment of this system becomes questionable when the zero-point contribution to the total moment of inertia is larger in magnitude than the classical one. We argue below that this in fact never occurs. 

The integral representation of \equ{CasE} for the Casimir energy with a $\delta$-distribution potential in the co-rotating frame gives an equally explicit expression for the zero-point contribution to the total angular momentum $\ell$,
\begin{align}
\label{ellofOmega}
\ell_\text{ZP}(\beta,\lambda )&=-\frac{\partial \E_c}{\partial\Omega}-I\Omega= -\hbar\int_0^\infty \xi d\xi\times\\
&\hspace{-8em}\times \frac {\xi(1\!-\!\beta^2)\!\left( (1\! +\!
           \beta)^2 e^{-2\pi\xi \beta}\! -\! (1\! -\! 
          \beta)^2  e^{2\pi\xi\beta}\!-\!4\beta e^{- 2\pi\xi} \right)\!+\! 2\beta \lambda R^2 e^{-2\pi\xi}} 
          {\lambda R^2\sinh(2\pi\xi)+ 
   4\xi (1-\beta^2) \sinh(\pi\xi(1 +\beta)) \sinh(\pi\xi (1 - \beta))}\ . \nonumber      
\end{align}
\begin{figure}
\includegraphics[width=3.5in]{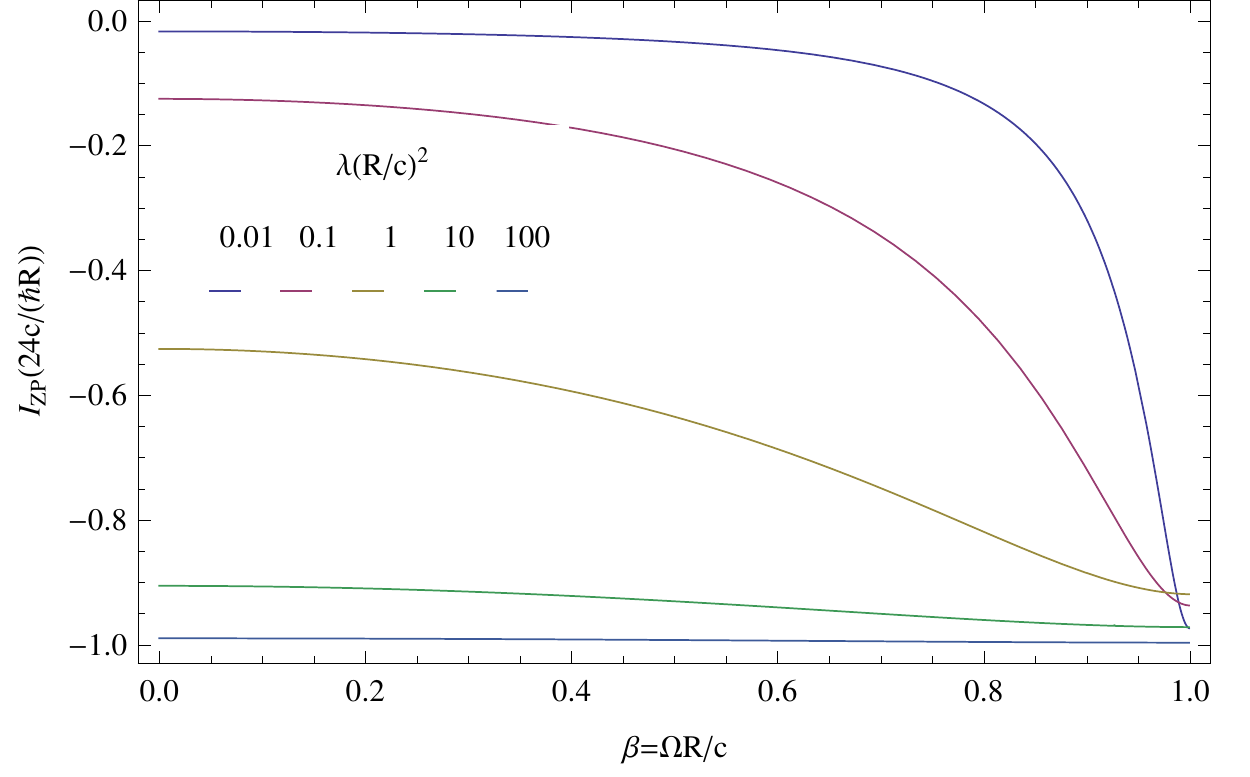}
\caption{ The zero-point contribution to the moment of inertia $I_{ZP}$ as a function of the dimensionless angular rotation speed $\beta=\Omega R/c$ for several dimensionless couplings $\lambda R^2/c^2$ of the $\delta$-distribution potential.  Note that $I_{ZP}$ becomes independent of the angular frequency only for $\lambda\sim\infty$, that is, for a Dirichlet condition. At finite coupling there is some \lq\lq{}slippage\rq\rq{} and the zero-point contribution to the moment of inertia is reduced in magnitude.    Note that $I_{ZP}$ is a bounded and monotonically decreasing function of $\beta$.}
\label{figI}
\end{figure}%

Although the zero-point contribution to the moment of inertia $I_{ZP}=\partial\ell_{ZP}/\partial\Omega$ is too lengthy to be reproduced here, one can deduce from \equ{ellofOmega} that $I_{ZP}$ is always negative and a monotonically decreasing function of the angular rotation speed $|\beta|<1$ (see \fig{figI}). It thus is bounded below by its value when the ring is rotating at the speed of light. From \equ{ellofOmega} one then finds,
\begin{align}
\label{IZP}
0&>I_{ZP}(R,\beta^2,\lambda)>I_{ZP}(R,1,\lambda)=\\
&=-\frac{\hbar R}{24 c}\left(1-\int_0^\infty\hspace{-.5em} \frac{6\lambda  (R/c)^2 \xi^2d\xi e^{-\pi \xi}}{(\lambda (R/c)^2+2\xi-2\xi e^{-\pi\xi})^2}\right)\ge-\frac{\hbar R}{24 c}\ ,\nonumber
\end{align}  
for any coupling $\lambda\ge 0$.

%\section{Conclusions}
Somewhat unexpected perhaps, the zero-point contribution to the total angular momentum in \equ{ellofOmega} also is bounded  in the accessible region  $|\beta|\leq 1$ by,
\bel{bound}
|\ell_\text{ZP}(\beta)|\leq |\ell_\text{ZP}(1)|\!=\!\frac{ \hbar}{24}\!\left(\!1\!-\!\!\int_0^\infty\hspace{-.8em}\frac{12\xi^2 d\xi}{(\frac{\lambda R^2}{c^2}+2\xi) e^{\pi\xi}-2\xi}\!\right)\!\ll \hbar,
\ee   
for any coupling $\lambda\ge 0$ and radius $R$. 

Since the smallest observable change in the total angular momentum in this quantum system is $\pm\hbar$, the upper bound of \equ{bound} together with \equ{ellofOmega} imply,
\bel{imp}
I\frac{\Delta\Omega}{\Delta\ell}=1-\frac{\Delta\ell_{ZP}}{\Delta\ell}\ge (1-1/24)>0\ .
\ee
We therefore have that either $\frac{\Delta\ell}{\Delta\Omega}=I_\text{tot}>0$ or that the moment of inertia $I$ of the collective coordinate is itself negative. Since the latter case would contradict the model  assumptions, we arrive at the conclusion that the total effective moment of inertia of the simplest Chernodub-device is always positive and its ground state is non-rotating once the classical contribution to the total angular momentum of the device is included. This classical contribution is necessary to conserve the total angular momentum of this semi-classical model.

Noting that only the total moment of inertia of the entire device can be measured  and not the contribution from quantum fluctuations alone, one can always renormalize and decompose the total moment of inertia of the device as,
\bel{renorm}
I_\text{tot}(\beta^2)=I+I_{ZP}(\beta^2) =I_\text{tot}(\beta_0^2)+(I_{ZP}(\beta^2)-I_{ZP}(\beta^2_0))\ ,
\ee
where $\beta_0$ is a reference rotation speed. If we choose $\beta_0=1$, the second term in \equ{renorm} is positive for all $\beta$ due to the lower bound of \equ{IZP}. Whether or not the total moment of inertia of the device is negative therefore depends exclusively on phenomenological input and can only be determined by a \emph{measurement}.  The renormalized form of \equ{renorm} pays tribute to the fact that the quantum fluctuations are not the whole story and also makes sense when $I_{ZP}(\beta^2_0)\rightarrow\infty$ but differences remain finite. Note that the negative contribution from quantum fluctuations in this model is irrelevant in \equ{renorm}.  The conclusion could be different only if quantum corrections to the total moment of inertia were unbounded below-- in this case $I_\text{tot}(\beta^2)$ invariably turns negative for \emph{some} value of $|\beta|<1$ and measuring $I_\text{tot}(\beta_0^2)$ determines only at which rotation speed this occurs.   

This simple and transparent model thus demonstrates that a negative zero-point contribution to the moment of inertia does not imply that the ground state of the complete quantum system could be self-rotating -- much as negative contributions to the mass from quantum fluctuations do not imply the existence of tachyons.

Note further that due to the relation in \equ{defOm}, a self-rotating ground state would imply that $\Omega(\bar\ell)=0$ at some finite value $\bar\ell\neq 0$. Assuming that $\Omega(\ell=0)=0$ as well, $\E_c(\Omega\sim 0)$ would have to be multi-valued at $\Omega=0$. This is not the case for a quadratic Hamiltonian such as $H_c$ with a unique ground state. 

The collective coordinate allows one to relate the Casimir energy in the stationary system to the one in the co-rotating frame by a Legendre transformation. Since $H_c$ of \equ{Hc} and the total angular momentum $\ell$ are commuting hermitian operators, one  therefore can conclude that the Casimir energies of the co-rotating and stationary frame are both real and that total angular momentum is conserved. No vacuum friction slows the rotation of this device. There is no transfer of angular momentum between the zero-point and classical contributions to the total angular momentum of this device. 

%\section{acknowledgments}
I would like to thank M. Chernodub for extensive discussions clarifying his model and for drawing my attention to the footnote in\cite{Landau:1980li}. This work was partly supported by the National Science Foundation with Grant no.~PHY0902054.

%-----------------------------------------------------
\bibliography{biblio/bChernodub}

%merlin.mbs apsrev4-1.bst 2010-07-25 4.21a (PWD, AO, DPC) hacked
%Control: key (0)
%Control: author (0) dotless jnrlst
%Control: editor formatted (1) identically to author
%Control: production of article title (0) allowed
%Control: page (1) range
%Control: year (0) verbatim
%Control: production of eprint (0) enabled
\begin{thebibliography}{12}%
\makeatletter
\providecommand \@ifxundefined [1]{%
 \@ifx{#1\undefined}
}%
\providecommand \@ifnum [1]{%
 \ifnum #1\expandafter \@firstoftwo
 \else \expandafter \@secondoftwo
 \fi
}%
\providecommand \@ifx [1]{%
 \ifx #1\expandafter \@firstoftwo
 \else \expandafter \@secondoftwo
 \fi
}%
\providecommand \natexlab [1]{#1}%
\providecommand \enquote  [1]{``#1''}%
\providecommand \bibnamefont  [1]{#1}%
\providecommand \bibfnamefont [1]{#1}%
\providecommand \citenamefont [1]{#1}%
\providecommand \href@noop [0]{\@secondoftwo}%
\providecommand \href [0]{\begingroup \@sanitize@url \@href}%
\providecommand \@href[1]{\@@startlink{#1}\@@href}%
\providecommand \@@href[1]{\endgroup#1\@@endlink}%
\providecommand \@sanitize@url [0]{\catcode `\\12\catcode `\$12\catcode
  `\&12\catcode `\#12\catcode `\^12\catcode `\_12\catcode `\%12\relax}%
\providecommand \@@startlink[1]{}%
\providecommand \@@endlink[0]{}%
\providecommand \url  [0]{\begingroup\@sanitize@url \@url }%
\providecommand \@url [1]{\endgroup\@href {#1}{\urlprefix }}%
\providecommand \urlprefix  [0]{URL }%
\providecommand \Eprint [0]{\href }%
\providecommand \doibase [0]{http://dx.doi.org/}%
\providecommand \selectlanguage [0]{\@gobble}%
\providecommand \bibinfo  [0]{\@secondoftwo}%
\providecommand \bibfield  [0]{\@secondoftwo}%
\providecommand \translation [1]{[#1]}%
\providecommand \BibitemOpen [0]{}%
\providecommand \bibitemStop [0]{}%
\providecommand \bibitemNoStop [0]{.\EOS\space}%
\providecommand \EOS [0]{\spacefactor3000\relax}%
\providecommand \BibitemShut  [1]{\csname bibitem#1\endcsname}%
\let\auto@bib@innerbib\@empty
%</preamble>
\bibitem [{\citenamefont {Chernodub}(2012{\natexlab{a}})}]{Chernodub:2012ry}%
  \BibitemOpen
  \bibfield  {author} {\bibinfo {author} {\bibfnamefont {M.N.}\ \bibnamefont
  {Chernodub}},\ }\bibfield  {title} {\enquote {\bibinfo {title} {{Permanently
  rotating devices: extracting rotation from quantum vacuum fluctuations?}}}\
  }\href@noop {} {\  (\bibinfo {year} {2012}{\natexlab{a}})},\ \Eprint
  {http://arxiv.org/abs/1203.6588} {arXiv:1203.6588 [quant-ph]} \BibitemShut
  {NoStop}%
%%CITATION = ARXIV:1203.6588;%%
\bibitem [{\citenamefont {Chernodub}(2012{\natexlab{b}})}]{Chernodub:2012em}%
  \BibitemOpen
  \bibfield  {author} {\bibinfo {author} {\bibfnamefont {M.N.}\ \bibnamefont
  {Chernodub}},\ }\bibfield  {title} {\enquote {\bibinfo {title} {{Rotating
  Casimir systems: magnetic-field-enhanced perpetual motion, possible
  realization in doped nanotubes, and laws of thermodynamics}},}\ }\href@noop
  {} {\  (\bibinfo {year} {2012}{\natexlab{b}})},\ \Eprint
  {http://arxiv.org/abs/1207.3052} {arXiv:1207.3052 [quant-ph]} \BibitemShut
  {NoStop}%
%%CITATION = ARXIV:1207.3052;%%
\bibitem [{\citenamefont {Shapere}\ and\ \citenamefont
  {Wilczek}(2012)}]{Shapere:2012nq}%
  \BibitemOpen
  \bibfield  {author} {\bibinfo {author} {\bibfnamefont {Alfred}\ \bibnamefont
  {Shapere}}\ and\ \bibinfo {author} {\bibfnamefont {Frank}\ \bibnamefont
  {Wilczek}},\ }\bibfield  {title} {\enquote {\bibinfo {title} {{Classical Time
  Crystals}},}\ }\href {\doibase 10.1103/PhysRevLett.109.160402} {\bibfield
  {journal} {\bibinfo  {journal} {Phys.Rev.Lett.}\ }\textbf {\bibinfo {volume}
  {109}},\ \bibinfo {pages} {160402} (\bibinfo {year} {2012})},\ \Eprint
  {http://arxiv.org/abs/1202.2537} {arXiv:1202.2537 [cond-mat.other]}
  \BibitemShut {NoStop}%
%%CITATION = ARXIV:1202.2537;%%
\bibitem [{\citenamefont {Wilczek}(2012)}]{Wilczek:2012tc}%
  \BibitemOpen
  \bibfield  {author} {\bibinfo {author} {\bibfnamefont {F.}~\bibnamefont
  {Wilczek}},\ }\bibfield  {title} {\enquote {\bibinfo {title} {{Quantum Time
  Crystals}},}\ }\href@noop {} {\  (\bibinfo {year} {2012})},\ \Eprint
  {http://arxiv.org/abs/1202.2539} {arXiv:1202.2539 [quant-ph]} \BibitemShut
  {NoStop}%
\bibitem [{\citenamefont {Bohr}\ and\ \citenamefont
  {Mottelson}(1969)}]{Bohr:1969ws}%
  \BibitemOpen
  \bibfield  {author} {\bibinfo {author} {\bibfnamefont {Aarge}\ \bibnamefont
  {Bohr}}\ and\ \bibinfo {author} {\bibfnamefont {Ben~R.}\ \bibnamefont
  {Mottelson}},\ }\href@noop {} {\emph {\bibinfo {title} {Nuclear Structure:
  Volume I: Single-Particle Motion; Volume II: Nuclear Deformations}}}\
  (\bibinfo  {publisher} {New York, W. A. Benjamin},\ \bibinfo {year}
  {1969})\BibitemShut {NoStop}%
\bibitem [{\citenamefont {Mottelson}(1976)}]{Mottelson:1976sc}%
  \BibitemOpen
  \bibfield  {author} {\bibinfo {author} {\bibfnamefont {Ben~R.}\ \bibnamefont
  {Mottelson}},\ }\bibfield  {title} {\enquote {\bibinfo {title} {{Elementary
  Modes of Excitation in the Nucleus}},}\ }\href@noop {} {\bibfield  {journal}
  {\bibinfo  {journal} {Science}\ }\textbf {\bibinfo {volume} {193}},\ \bibinfo
  {pages} {287--294} (\bibinfo {year} {1976})}\BibitemShut {NoStop}%
\bibitem [{\citenamefont {Sutcliffe}(1993)}]{Sutcliffe:1993su}%
  \BibitemOpen
  \bibfield  {author} {\bibinfo {author} {\bibfnamefont {Paul~M.}\ \bibnamefont
  {Sutcliffe}},\ }\bibfield  {title} {\enquote {\bibinfo {title} {Classical and
  quantum kink scattering},}\ }\href {\doibase 10.1016/0550-3213(93)90243-I}
  {\bibfield  {journal} {\bibinfo  {journal} {Nuclear Physics B}\ }\textbf
  {\bibinfo {volume} {393}},\ \bibinfo {pages} {211 -- 224} (\bibinfo {year}
  {1993})}\BibitemShut {NoStop}%
\bibitem [{\citenamefont {Graham}\ \emph {et~al.}(2002)\citenamefont {Graham},
  \citenamefont {Jaffe},\ and\ \citenamefont {Weigel}}]{Graham:2002fi}%
  \BibitemOpen
  \bibfield  {author} {\bibinfo {author} {\bibfnamefont {Noah}\ \bibnamefont
  {Graham}}, \bibinfo {author} {\bibfnamefont {Robert~L.}\ \bibnamefont
  {Jaffe}}, \ and\ \bibinfo {author} {\bibfnamefont {Herbert}\ \bibnamefont
  {Weigel}},\ }\bibfield  {title} {\enquote {\bibinfo {title} {{Casimir effects
  in renormalizable quantum field theories}},}\ }\href {\doibase
  10.1142/S0217751X02010224} {\bibfield  {journal} {\bibinfo  {journal}
  {Int.J.Mod.Phys.}\ }\textbf {\bibinfo {volume} {A17}},\ \bibinfo {pages}
  {846--869} (\bibinfo {year} {2002})},\ \Eprint
  {http://arxiv.org/abs/hep-th/0201148} {arXiv:hep-th/0201148 [hep-th]}
  \BibitemShut {NoStop}%
\bibitem [{\citenamefont {Landau}\ and\ \citenamefont
  {Lifshitz}(1980)}]{Landau:1980li}%
  \BibitemOpen
  \bibfield  {author} {\bibinfo {author} {\bibfnamefont {L.~D.}\ \bibnamefont
  {Landau}}\ and\ \bibinfo {author} {\bibfnamefont {E.~M.}\ \bibnamefont
  {Lifshitz}},\ }\href@noop {} {\emph {\bibinfo {title} {Statistical Physics,
  3rd ed.}}}\ (\bibinfo  {publisher} {Oxford, Pergamon},\ \bibinfo {year}
  {1980})\BibitemShut {NoStop}%
\bibitem [{\citenamefont {Bloch}(1929)}]{Bloch:1929bl}%
  \BibitemOpen
  \bibfield  {author} {\bibinfo {author} {\bibfnamefont {Felix}\ \bibnamefont
  {Bloch}},\ }\bibfield  {title} {\enquote {\bibinfo {title} {{\"{U}}ber die
  {Q}uantenmechanik der {E}lektronen in {K}ristallgittern},}\ }\href
  {http://dx.doi.org/10.1007/BF01339455} {\bibfield  {journal} {\bibinfo
  {journal} {Zeitschrift f{\"u}r Physik A: Hadrons and Nuclei}\ }\textbf
  {\bibinfo {volume} {52}},\ \bibinfo {pages} {555--600} (\bibinfo {year}
  {1929})}\BibitemShut {NoStop}%
\bibitem [{\citenamefont {Kirsten}(2001)}]{Kirsten:2001wz}%
  \BibitemOpen
  \bibfield  {author} {\bibinfo {author} {\bibfnamefont {Klaus}\ \bibnamefont
  {Kirsten}},\ }\bibfield  {title} {\enquote {\bibinfo {title} {{Spectral
  functions in mathematics and physics}},}\ }\href@noop {} {\  (\bibinfo {year}
  {2001})}\BibitemShut {NoStop}%
%%CITATION = INSPIRE-567159;%%
\bibitem [{\citenamefont {Kampen}\ \emph {et~al.}(1968)\citenamefont {Kampen},
  \citenamefont {Nijboer},\ and\ \citenamefont {Schram}}]{VanKampen:1968}%
  \BibitemOpen
  \bibfield  {author} {\bibinfo {author} {\bibfnamefont {N.G.~Van}\
  \bibnamefont {Kampen}}, \bibinfo {author} {\bibfnamefont {B.R.A.}\
  \bibnamefont {Nijboer}}, \ and\ \bibinfo {author} {\bibfnamefont
  {K.}~\bibnamefont {Schram}},\ }\bibfield  {title} {\enquote {\bibinfo {title}
  {On the macroscopic theory of {V}an der {W}aals forces},}\ }\href {\doibase
  10.1016/0375-9601(68)90665-8} {\bibfield  {journal} {\bibinfo  {journal}
  {Physics Letters A}\ }\textbf {\bibinfo {volume} {26}},\ \bibinfo {pages}
  {307 -- 308} (\bibinfo {year} {1968})}\BibitemShut {NoStop}%
\end{thebibliography}%
%\nocite{*} %%% Will print the complete bib data.
%-----------------------------------------------------

\end{document}